\documentclass[aps,pre,showpacs,twocolumn,amsmath,amssymb,superscriptaddress]{revtex4-1}
\usepackage{graphicx,bbm,color}

\newcommand{\be}{\begin{equation}}
\newcommand{\ee}{\end{equation}}

\newcommand{\bra}[1]{{\langle #1 \vert}}

\newcommand{\ket}[1]{{\vert #1 \rangle}}

\newcommand{\ave}[1]{{\langle #1\rangle}}
\newcommand{\ii}{ {\rm i} }
\newcommand{\dd}{ {\rm d} }
\newcommand{\ZZ}{\mathbb{Z}}

\newcommand{\CC}{\mathbb{C}}

\newcommand{\z}{{\rm z}}

\newcommand{\mm}[1]{{\mathbf{#1}}}
\newcommand{\half}{\frac{1}{2}}
\newcommand{\ua}{\uparrow}
\newcommand{\da}{\downarrow}

\def\tr{{{\rm tr}}}
\def\ad{{\,{\rm ad}\,}}
\def\one{\mathbbm{1}}

\def\DD{\hat{\cal D}}

\newcommand{\LL}{{\hat{\cal L}}}

\begin{document}

\title{Charge and spin current statistics of the open Hubbard model with weak coupling to the environment} 
\author{Berislav Bu\v ca}
\affiliation{Department of Medical Physics and Biophysics, University of Split School of Medicine, 21000 Split, Croatia}
\author{Toma\v z Prosen}
\affiliation{Department of Physics, Faculty of Mathematics and Physics, University of Ljubljana, Jadranska 19, 1000 Ljubljana, Slovenia}

\begin{abstract}
Based on generalization and extension of previous work  [Phys. Rev. Lett. {\bf 112}, 067201 (2014)] to multiple independent markovian baths we will compute the charge and spin current statistics of the open Hubbard model with weak system-bath coupling up to next-to-leading order in the coupling parameter. The physical results are related to those for the $XXZ$ model in the analogous setup implying a certain universality which potentially holds in this class of nonequilibrium models. 
\end{abstract}

 
\maketitle
\section{Introduction}
\label{sec:intro}

One of the key challenges of modern physics is understanding non-equilibrium behavior of both quantum and classical system \cite{EisertFriesdorfGogolin}. In contrast to ordinary equilibrium statistical physics many approaches for understanding non-equilibrium physics exist, as there are many ways to drive a system from equilibrium. From quenches to open quantum systems one is interested often in similar concepts, such as transport of a certain quantity, that is the behavior of current of this quantity. The standard approach is to compute the expectation value of an observable (typically the current in the case of transport) with respect to some state of the system. In open quantum systems \cite{openbook} this is usually the expectation value in the non-equilibrium steady state (NESS) to which the system evolves can evolve to in the infinite time limit. However, this approach is not the whole story and one may gain significantly more physical insights by studying the whole probability distribution of the observable. This is usually more difficult than computing just the expectation value.

This approach is additionally complicated in the quantum setting as measurements of an observable (such as the current) influence the time evolution of the system. One possible resolution to this issue is to look at a framework in which the non-equilibrium driving (via coupling to the baths) effectively performs the measurement as well. The framework of Markovian quantum master equations (Lindblad equation) is particularly well-suited for setting such a protocol. The study of transport in this context has attracted a lot of attention recently. For some very recent examples of such studies see Refs.~\cite{Arenas1, Ilievski, EverestLesanovsky,ZnidScarVar,ManzanoChuangCao,KPS2,IlievskiZunkovic,Prosen2, WolffSheikhanKollath} and a recent review \cite{Prosenreview} in the context of exact solutions. 

The Lindblad master equation \cite{Lindblad} for the system's density operator $\rho(t)$ can be written as
\be
\frac{\dd \rho(t)}{\dd t} = -\ii [H,\rho(t)] + \varepsilon \left ( \DD^{\rm jump} \rho(t) + \DD^{\rm diss}\rho(t) \right ),
\label{eq:lindblad}
\ee
where $H$ is the Hamiltonian of the system generating the unitary part  of the time evolution, and the dissipator $\DD$ implements the action of environment on the system. The latter is split into {\em the quantum jumps} and {\em the dissipation}
\be
\DD^{\rm jump}\rho :=\sum_\mu L_m \rho L^\dagger_m,\quad
\DD^{\rm diss} \rho := \frac{1}{2}\sum_m \{ L^\dagger_m L_m,\rho\},
\ee
where, as in the previous work \cite{Prosen1, countstatweak}, the system-bath coupling parameter $\varepsilon$ is assumed to be small enough to allow the use of perturbation theory. The applicability of this method, namely the convergence radius of the perturbation series in this parameter is discussed in \cite{LemosProsen},
where the authors point out issues of exponentially (in system size) decreasing size of the radius of convergence when one does perturbation for small $\varepsilon$. Thus, the methods relying on perturbation expansion in $\varepsilon$ cannot be applied in the thermodynamic limit, but can still be useful for finite system sizes. 

Full counting statistics is the method which allows us to formulate in an elegant and efficient way the computation of cumulants of a current of a conserved quantity \cite{scully, levitov, rev}, provided that certain conditions, elaborated in the next section hold. For some very recent results, as well as the related large deviation theory, in the open quantum framework in this area see, e.g., Refs.~\cite{Garrahan, ZnidaricFCS,ZnidaricFCS2,ZnidaricFCS3,Monthus,PigeonXuereb,ManzanoHurtado,MK, flucthydro}.

In this paper we will study the multi-bath generalization of the results obtained in Ref.~\cite{countstatweak}. Namely, we will show how a simple generalization of the full counting statistics for spin-1/2 systems weakly coupled to multiple baths can be achieved in the leading order in $\varepsilon$. 

Afterwards, we will give a solution to the full counting statistics in next-to-leading (third) order for the open 1D (fermi) Hubbard model. The Hubbard model is a paradigm of many-body quantum physics. It describes numerous fundamental transport phenomena, in particular its two dimensional version is believed to capture superconductivity in cuprates. The 1D Hubbard model is integrable and exactly solvable through the Bethe ansatz \cite{BaxterBook, EsslerBook, LiebWu}.  By mapping the fermionic Hubbard model to a spin ladder we will find the full counting statistics of both the spin and charge currents in the third order of system-bath coupling. 

The solution is based on the construction \cite{ProsenHubbard} of the exact steady state density operator for the open maximally driven Hubbard model, which was later put into a Lax form in Ref.~\cite{PopkovProsen}.  We show how to generate the solution to our counting problem using this infinite dimensional Lax operator. The solution, much like the previously discussed one for the $XXZ$ spin chain, is given by almost conserved charges which break the $\ZZ_2$ symmetry of the model (in this case the particle-hole symmetry) \cite{ProsenHubbardCharge}. 

\section{Full counting statistics of open systems with multiple baths}
\label{sec:multibaths}

We aim at computing the full spin or charge exchange (and current) statistics in the limit of weak system-bath coupling when the Markovian approximation for the evolution of system's density matrix $\rho(t)$ is appropriate \cite{openbook}.
In contrast to previous work \cite{countstatweak} we will consider multiple, $D$ species of spins-1/2 labeled by a superscript $d= 1, 2 \ldots D$ associated with the physical site $j$, i.e., $\sigma^{d, \pm}_j$, $\sigma^{d, \z}_j$, where these are the standard Pauli operators acting on a tensor product space $(\CC^2)^{\otimes D n}$. Also, assume that our system is coupled to several sets of baths, i.e., in contrast to the previous work studied in \cite{countstatweak} we allow for the system to be coupled to more than two baths. 
Let $H$ conserve one or more extensive observables with integer spectrum $M_d$, e.g. magnetization, charge, particle number etc, $[H,M_d]=0$. The subscript $d$ in $M_d$ denotes that $M_d$ is composed solely of spin operators of species $d$. 

The Lindblad jump operators composing the dissipator, which are assumed to act ultra-locally (on one spin site only), represent the action of each bath on the system. For every spin species $d$ there exists one pair of baths corresponding to and driving the current of that spin species. We will then consider the flow (current) of observable $M_d$ from a {\em subset} of baths ${\cal I}$ (with the jump operator composed of one element in the pair of spin operators of species $d$ for every $d$, say $\sigma^{d,+}$)  to its complement ${\cal O}$  (with the jump operator composed of the other element in the pair of spin operators of species $d$ for every $d$, say $\sigma^{d,-}$). 
We write $\tau_\mu:=1$ for $\mu\in{\cal I}$, and $\tau_\mu:=-1$ for $\mu\in{\cal O}$. 
Let us denote the amount of quantity $M_d$, say $\sum_j \sigma^{d,\z}_j$, transported in time $t$ from ${\cal I}$ to ${\cal O}$ by $N_d(t)$. In the open system's framework, $N_d(t)$ is exactly the sum of $\delta_\mu$ for all the jumps for $\mu\in{\cal I}$ minus the sum of $\delta_\mu$ for all the jumps of $\mu\in{\cal O}$.
 In other words, we let the jump operators change $M_d$ by $\delta_\mu\in\ZZ$, and if they drive current of $M_d$ from ${\cal I}$ to ${\cal O}$ we say that the current is driven in the positive direction ($\tau_\mu=1$) and if the drive current of $M_d$ in the other direction then we say that the current is driven in the negative direction ($\tau_\mu=-1$). 


In order to apply the counting field formalism we introduce now a counting vector field  $\vec{\chi}=\{\chi_1, \chi_2\,\ldots, \chi_d, \ldots\}$
where ${\chi}_d$ counts, for spin species $d$, how much of charge $M_d$ was transferred during the time evolution of the system. 
For simplicity we will take $D$ \emph{symmetric} current driving dissipators acting on "complementary" sites of spin species $d$, each pair of sites being labeled by $m_\nu$ and each site in the pair by $\nu=\pm$. Then we may split the dissipator in the following manner,
\begin{eqnarray}
\DD^{\rm jump}_{\vec{\chi}}&:=& \sum_{m,d} (1+\mu^d_m) \DD^{\rm jump}_{d, m,{+}}+(1-\mu^d_m) \DD^{\rm jump}_{d, m,{-}},\\
\DD^{\rm diss}&:=& \sum_{m,d} (1+\mu^d_m) \DD^{\rm diss}_{d, m,{+}}+(1-\mu^d_m) \DD^{\rm diss}_{d, m,{-}}, 
\end{eqnarray}
where,
\begin{eqnarray}
\DD^{\rm jump}_{d, m,{+}} \rho &:=& e^{\ii \chi_d} \left ( L_{m_+}^{d,+}\rho (L_{m_+}^{d,+})^\dagger+ L_{m_-}^{d,-} \rho (L_{m_-}^{d,-})^\dagger \right ), \label{eq:countdiss} \\
\DD^{\rm jump}_{d, m,{-}} \rho &:=& e^{-\ii \chi_d} \left ( L_{m_-}^{d,+} \rho (L_{m_-}^{d,+})^\dagger+ L_{m_+}^{d,-}\rho (L_{m_+}^{d,-})^\dagger \right ), \nonumber
\end{eqnarray}
and,
\begin{eqnarray}
\DD^{\rm diss}_{d, m,{+}}\rho&:=&\frac{1}{2}\left (\{ (L_{m_+}^{d, +})^\dagger L_{m_+}^{d, +},\rho\}+ \{(L_{m_-}^{d, -})^\dagger  L_{m_-}^{d, -},\rho\}  \right )\\
\DD^{\rm diss}_{d, m,{-}}\rho&:=&\frac{1}{2} \left( \{ ( L_{m_-}^{d, +})^\dagger L_{m_-}^{d, +} ,\rho\} +\{(L_{m_+}^{d, -})^\dagger L_{m_+}^{d, -} ,\rho\} \right), \nonumber
\end{eqnarray}
where the index $m$ corresponds to the pair of sites on which the dissipator acts, the sub-subscript, $\mu=\pm$ in $m_\mu$, denotes on which site in the pair of sites $m$ a Lindblad jump operator, $L_{m_\mu}^{d, \nu}=\sigma^{d, \nu}_{m_\mu}$, acts and $\nu=\pm$ in $L_{m_\mu}^{d, \nu}=\sigma^{d, \nu}_{m_\mu}$, denotes either a raising or lowering operator 
\begin{equation}
[M_d,L_{m_{\mu}}^{d, \nu}] = \nu L_{m_{\mu}}^{d, \nu}
\end{equation} 
and $d$ is the species of spin on which the Lindblad jump operator acts.  The sign of the product $\mu \nu= \pm$ determines whether a Lindblad operator $L_{m_\mu}^{d, \nu}$ drives the current of $M_d$ in the positive or negative direction, i.e., $\tau_\mu=1$ or $\tau_\mu=-1$, respectively. 

With this modification the Lindblad equation \eqref{eq:lindblad} is deformed in the following way, 
\be
\frac{\dd \rho(t)}{\dd t} = \LL_{\vec{\chi}} \rho(t) := -\ii [H,\rho(t)] + \varepsilon \left ( \DD^{\rm jump}_{\vec{\chi}} \rho(t) + \DD^{\rm diss}\rho(t) \right ).
\label{eq:lindblad1}
\ee
The eigenvalue problem for the leading (with maximal real part) eigenvalue $\lambda(\vec{\chi})$ of this deformed Liouvillian $\LL_{\vec{\chi}}$ is given as, 
\be
\left [-\ii \ad H + \varepsilon \left ( \DD^{\rm jump}_{\vec{\chi}} + \DD^{\rm diss} \right ) \right ]\rho(\vec{\chi}) = \lambda(\vec{\chi})\rho(\vec{\chi}), 
\label{eq:gen}
\ee
where $\rho(\vec{\chi})$ is the corresponding right eigenvectors, which will be assumed to be unique (as per the conditions of the Evans theorem \cite{Evans}). We have also defined the superoperator $\ad H x:= [H, x]$ which acts on the space of operators. Note that  $\rho(\vec{\chi}=0)$ corresponds to the non-equilibrium steady state of the undeformed Lindblad equation \eqref{eq:lindblad}. We remark that the steady state can be also alternatively degenerate and symmetry reducible \cite{sym1, sym2, sym3}. Such a case is studied in Ref.~\cite{ManzanoHurtado}. 

The leading eigenvalue $\lambda(\vec{\chi})$ in \eqref{eq:gen} will be the cumulant generating function for the currents $I_d$ of $M_d$,
\be
\ave{I^m_d}_c := \lim_{t \to \infty}\frac{1} {t} \ave{[N_d(t)]^m}_c =  \frac {\partial^{m} \lambda(\vec{\chi}) }{\partial (\ii \chi_d)^{m}} \Big |_{\vec{\chi} \to 0}. \label{cs0_hub}
\ee
To give an intuitive picture of why this method works we will now give a short elaboration for only one spin species $D=1$, writing $\vec{\chi}\equiv \chi$.  The method of full counting statistics can be understood most easily by observing that since $H$ commutes with the quantity of interest, and the Lindblad operators change the value of this quantity by well-defined (classical) amounts one may project the density matrix of the system $\rho(t)$ on a subspace of $N$ spin transfers between the pair of baths in time $t$ which we denote as $\rho_N(t)$. Thus, the probability of $N$ spin transfers in time $t$ is given as $P_N(t)=\tr \rho_N(t)$. We can then perform a Fourier transform in $N$ and then find that the dissipator in the original Lindblad equation \eqref{eq:lindblad} is modified to include the counting field in the form written in Eq.~\eqref{eq:countdiss}. Then in the long-time limit where $\rho(\chi, t) \approx e^{\lambda(\chi) t} \rho(\chi,t= 0) $, the leading (with largest real part) eigenvalue $\lambda(\chi)$ is the cumulant generating function for the current \eqref{cs0_hub}.

The perturbation expansion is given by 
 \be
\rho (\vec{\chi}) = \sum_{p=0}^\infty (\ii \varepsilon)^p \rho^{(p)}, \quad \lambda (\vec{\chi})  = \sum_{p=1}^\infty \varepsilon^{2p-1} \lambda^{(2p-1)}. \label{eq:pert}
\ee
The first two orders obtained by plugging in \eqref{eq:pert} in are,
\begin{align} 
(\!\ad H) \rho^{(0)} &=0, \label{zo} \\ 
(\!\ad H) \rho^{(1)} + \hat{\cal D}_{\vec{\chi}} \rho^{(0)}  &= \lambda^{(1)} \rho^{(0)}, \label{fo}
\end{align}
where we also defined the total dissipator as $\DD_{\vec{\chi}}:=\DD^{\rm jump}_{\vec{\chi}} + \DD^{\rm diss}$. In general any linear combination of operators which commute with $H$ satisfy the zeroth-order equation \eqref{zo}. So we may write in general this solution using a linearly independent combination of $Q_k$ such that $[H,Q_k]=0$,
\be
\rho^{(0)}=\sum_k  \alpha_k Q_k. \label{eq:rho0gen}
\ee
Note that since that we have assumed that the corresponding right eigenvector is unique not every possible solution to the zeroth-order equation \eqref{zo} can be the correct leading order in the perturbation expansion \eqref{eq:pert}. In order to find the correct one, we also demand \cite{Prosen1, countstatweak} that the solution to the first-order equation \eqref{fo} exists. 
To do this we will demand a generalization of the simple algebraic condition on the image space of $\ad H$ (see \cite{countstatweak}) to hold, 
\begin{equation}
-\sigma^{d, \z}_{m_+} + \sigma^{d, \z}_{m_ {-}} \in  {\rm Im}\ad H,
\label{eq:cond_hub}
\end{equation}
 for every spin species $d$ and for every pair of sites $m$ on which the Lindblad jump operators act. 
This requirement will be fulfilled if a generalization of the previous parity symmetry condition studied in \cite{countstatweak} holds. That is, an operator $P$ ($P^2=\one)$
should exist, which satisfies either
\begin{equation} 
 \quad P \sigma^{d, \z}_{m_+}  =  \sigma^{d, \z}_{m_ {-}} P,\quad {\rm or} \quad  P  \sigma^{d, \z}_{m_ {+, -}} = - \sigma^{d, \z}_{m_ {+, -}} P, \label{eq:Psym2_hub}
\end{equation}
and commutes with the Hamiltonian $[P,H]=0$ together with all the conserved operators $[P,Q_k]=0$. The proof of this statement is the same as in Ref.~\cite{countstatweak}: Since $\ad H$ is Hermitian its image is orthogonal w.r.t. Hilbert-Schmidt inner-product to all $Q_k$. Thus the condition \eqref{eq:cond_hub} is equivalent to: $\tr(-\sigma^{d, \z}_1+\sigma^{d, \z}_n)Q_k = 0, \forall Q_k$. From this we may directly see that $\tr(-\sigma^{d, \z}_{m_+} + \sigma^{d, \z}_{m_ {-}} )Q_k = \tr P(-\sigma^{d, \z}_{m_+} + \sigma^{d, \z}_{m_ {-}} )Q_k P =  -\tr (-\sigma^{d, \z}_{m_+} + \sigma^{d, \z}_{m_ {-}} )Q_k$. From which follows that  $\tr(-\sigma^{d, \z}_{m_+} + \sigma^{d, \z}_{m_ {-}} )Q_k=0$. 

It is then trivial to show that given,
\begin{equation}
\rho^{(0)}= \frac{1}{(2d)^n} \one, \label{zero_hub_sol}
\end{equation}
and,
\begin{equation}
  \lambda^{(1)}= \sum_d 2(-1+\cos( \chi_d) - \ii \mu_d \sin( \chi_d)), \label{zero_hub_eig}
\end{equation}
a solution to the first-order equation \eqref{fo} exists. This can be done by plugging the expressions \eqref{zero_hub_sol} and  \eqref{zero_hub_eig} into \eqref{fo} and showing that \eqref{eq:cond_hub} holds. 

By performing a suitable transformation $\vec{\chi} \to -\ii \vec{\chi}$ we switch to (provided there are no issues with analyticity) what is called large-deviation theory \cite{Oono, Touchette}. 

 \section[Charge and spin statistics of the open Hubbard model...]{Charge and spin statistics of the open Hubbard model at weak system-bath coupling}
 
The first-order result of the previous section was universal and only depended on the local nature of jump operators and existence of the parity symmetry. In this section we
show that one can compute higher order in $\varepsilon$ for a multispecies integrable model, namely the 1D Hubbard model.
The fermionic Hubbard model with $n$ sites is given by the following Hamiltonian,
\begin{equation}
H=-2\sum_{s,x} (c^\dagger_{s,j}c_{s,j+1}+c^\dagger_{s,j+1}c_{s,j} ) +  u \sum_{j} (2n_{\uparrow,j}-1)(2n_{\downarrow,j}-1) \nonumber\\
\label{eq:HubHfermi}
\end{equation}
where $c_{s,j}$ are the canonical Fermi operators on site $j$ and $j\in\{1\ldots n\}$, $s\in\{\ua,\da\}$ is the spin of each fermion, $u$ is a non-dimensional parameter determining the interaction strength, and $n_{s,j}=c^\dagger_{s,j} c_{s,j}$ is the local charge density. 

We now apply the Wigner-Jordan transformation, 
\begin{equation}
c_{\uparrow,j}=P^{(\sigma^1)}_{j-1} \sigma_j^{1,-},\quad c_{\downarrow,j}=P^{(\sigma^1)}_n P^{(\sigma^2)}_{j-1} \sigma_j^{2,-}
\end{equation}
where
$P^{(\sigma^1)}_j:=\sigma_1^{1, \rm z}\sigma_2^{1,\z} \cdots \sigma_j^{1, \rm z}$, $P^{(\sigma^2)}_j:=\sigma_1^{2, \rm z}\sigma_2^{2, \z} \cdots \sigma_j^{2, \rm z}$, and of course $\sigma^{1,0}_j\equiv \sigma^{2,0}_j\equiv\one$. 
Thus we transform the fermionic Hubbard model into a spin ladder, 
\begin{align}
&H = 2  \sum_{j=1}^{n-1} \sum_{d=1}^2 \left ( \sigma^{d,+}_j \sigma^{d,-}_{j+1} +  \sigma^{d,-}_{j} \sigma^{d,+}_{j+1} \right ) + \frac{u}{2}   \sum_{j=1}^{n} \sigma^{1, \rm{z}}_j\sigma^{2, \rm{z}}_j,  \label{eq:Hhub} \\
&h_{j,j+1}=2 \sum_{d=1}^2 \left ( \sigma^{d,+}_j \sigma^{d,-}_{j+1} +  \sigma^{d,-}_{j} \sigma^{d,+}_{j+1} \right ) \nonumber \\
&+ \frac{u}{2} \left (  \sigma^{1, \rm{z}}_j\sigma^{2, \rm{z}}_j+\sigma^{1, \rm{z}}_{j+1}\sigma^{2, \rm{z}}_{j+1} \right)  
\label{eq:Hhubloc1}
\end{align}
We will also likewise define two local "kinetic energy" densities for the two species of spin,
\begin{align}
&h^{\sigma^{1}}_{j,j+1}:=2 \sigma^{1,+}_j \sigma^{1,-}_{j+1} + 2 \sigma^{1,-}_{j} \sigma^{1,+}_{j+1},  \\
&h^{\sigma^2}_{j,j+1}:=2 \sigma^{2,+}_{j} \sigma^-_{2,j+1} + 2 \sigma^{2, -}_j \sigma^{2,+}_{j+1} 
\label{eq:Hhubloc}
\end{align}
Note that the Hubbard Hamiltonian conserves the numbers of spin-up and spin-down electrons, 
$M_{\sigma^1}=\sum_{j=1}^n \half(\sigma^{1, \z}_j + \one)$, $M_{\sigma^2} = \sum_{j=1}^n \half(\sigma^{2,\z}_x + \one)$, $[H,N_{\sigma^{1,2}}]=0$. 

We now continue for sake of being self-contain by quickly reviewing the results of Refs.~\cite{ProsenHubbard} and  \cite{PopkovProsen} on the open Hubbard model 
where it has been shown that there exists an infinite dimensional Lax operator seemingly unrelated to the known one discovered by Shastry \cite{Shastry}. 

All the following definitions are given as in Ref.~\cite{PopkovProsen}. Define  $\mm{S},\acute{\mm{S}},\grave{\mm{S}},\mm{T},\acute{\mm{T}},\grave{\mm{T}} \in {\rm End}({\cal H}_{\rm a}\otimes {\cal H}_{\rm p})$, and $\mm{X},\mm{Y}\in {\rm End}({\cal H}_{\rm a})$ (scalars over ${\cal H}_{\rm p}$), where ${\cal H}_{\rm a} $ is an infinite dimensional auxiliary space.  We use labelling the basis vectors of the auxiliary space as  ${\cal V}=\{0^{+},\frac{1}{2}^{+},\frac{1}{2}^{-},1^{-},1^{+},\frac{3}{2}^{+},\frac{3}{2}^{-},2^{-},2^{+}\ldots\}$ so that\cite{footnote}
 ${\cal H}_{\rm a}={\rm lsp}\{\ket{v}; v\in{\cal V}\}$. Introduce another definition, namely the components $ \mm{S}^s,\mm{T}^s \in {\rm End}({\cal H}_{\rm a})$, $\mm{S}=\sum_{s \in \{+,-0, \z\}} \mm{S}^s \sigma^{1, s}$, $\mm{T}=\sum_{s \in \{+,-0, \z\}} \mm{T}^s \sigma^{2, s}$ and likewise for $\acute{\mm{S}},\grave{\mm{S}},\acute{\mm{T}},\grave{\mm{T}}$. 

Then $\mm{S}$ is given by the following matrix representations, 
\begin{eqnarray}
&&\mm{S}^+ =\sqrt{2} \sum_{k=0}^\infty \left(\ket{k^+}\bra{k\!+\!\half^+} + \ket{k\!+\!\half^-}\bra{k\!+\!1^-}\right),\label{eq:Sp} \\
&&\mm{S}^-  = \sqrt{2} \sum_{k=0}^\infty (-1)^k \left(\ket{k\!+\!\half^+}\bra{k^+} + \ket{k\!+\!1^-}\bra{k\!+\!\half^-}\right),\nonumber\\
&&\mm{S}^0 = \sum_{k=0}^\infty \bigl(\ket{2k^+}\bra{2k^+} + \ket{2k\!+\!\half^+}\bra{2k\!+\!\half^+} \nonumber \\
&&\qquad+ \ket{2k\!+\!1^-}\bra{2k\!+\!1^-}+ \ket{2k\!+\!\half^-}\bra{2k\!+\!\half^-}\bigr)\nonumber\\
&&\quad+ \phi\sum_{k=1}^\infty \left( \ket{2k\!-\!\half^+}\bra{2k\!-\!\half^+} + \ket{2k^-}\bra{2k^-}\right), \nonumber \\
&&\mm{S}^\z = \sum_{k=1}^\infty \bigl(\ket{2k\!-\!1^+}\bra{2k\!-\!1^+} + \ket{2k\!-\!\half^+}\bra{2k\!-\!\half^+} \nonumber \\
&&\qquad + \ket{2k^-}\bra{2k^-} + \ket{2k\!+\!\half^-}\bra{2k\!+\!\half^-}\bigr)  \nonumber\\
&&\quad + \phi\sum_{k=0}^\infty\left(\ket{2k\!+\!\half^+}\bra{2k\!+\!\half^+} + \ket{2k\!+\!1^-}\bra{2k\!+\!1^-}\right), \nonumber
\end{eqnarray}
where $\phi\in\CC$ is a free parameter. Defining an operator $\mm{G}$ which interchanges spin species in physical space, i.e., $\mm{G} \sigma^{1, s} \mm{G} = \sigma^{2,s}$, $\mm{G} \sigma^{2, s} \mm{G} = \sigma^{1,s}$ and in the auxiliary space operates as $\mm{G}\ket{k^\pm} := \ket{k^\pm}$, $\mm{G} \ket{k\!+\!\half^\pm} := \ket{k\!+\!\half^\mp}$, $k \in \ZZ^+$, $\mm{T}$ is given by $\mm{T}=\mm{G} \mm{S} \mm{G}$. 

The operator $\mm{X}$ is given by,
\begin{eqnarray}
\mm{X} &=& \ket{0^+}\bra{0^+} +
 \sum_{k=1}^\infty (-1)^{k}\!\!\!\sum_{\nu,\nu'\in\{-,+\}}   \ket{k^{\nu}} X^{\nu,\nu'}_k  \bra{k^{\nu'}}  \label{eq:Xansatz}\\
&+& w \sum_{k=0}^\infty (-1)^k \left(\ket{k\!+\!\half^+}\bra{k\!+\!\half^+} + \nonumber
\ket{k\!+\!\half^-}\bra{k\!+\!\half^-}\right), 
\end{eqnarray}
where $X_k=\{ X^{\nu,\nu'}_k \}_{\nu,\nu'\in\{-,+\}}$ are $2\times2$ matrices 
\begin{equation}
X_k(\phi,w)=\begin{pmatrix}
-(w+k u)w & 1-(w+k u)w(1-\phi^2)\cr
 -k u w & 1- k u w (1-\phi^2),
\end{pmatrix}.
\label{Xk}
\end{equation}
and $w \in \CC$ is another free parameter. 
The operator $\mm{Y}$ is defined as,
\begin{equation}
\mm{Y}=-2\phi u\sum_{k=0}^{\infty}\bigl(\ket{k^+}\bra{k^+} + \ket{k\!+\!1^-}\bra{k\!+\!1^-}\bigr).
\end{equation} 
Demanding that $\mm{X}$ is invertible, that is $w\neq 0$, $\det X_k \neq 0$, we may define implicitly, 
\begin{eqnarray}
&&\acute{\mm{S}}^+\mm{X}  =  -2\sqrt{2}\sum_{k=1}^{\infty}(-1)^k X^{+-}_k \ket{k^-}\bra{k\!+\!\half^+}, \label{eq:Sb}\\
&&\acute{\mm{S}}^-\mm{X} =   -2\sqrt{2}\sum_{k=1}^{\infty} X^{-+}_k\ket{k^+}\bra{k\!-\!\half^-}, \nonumber\\
&&\mm{X}\grave{\mm{S}}^{+}  =  2\sqrt{2}\sum_{k=1}^{\infty}
(-1)^k X^{+-}_k\ket{k\!-\!\half^-}\bra{k^+} \nonumber\\
&&\mm{X}\grave{\mm{S}}^- =  -2\sqrt{2}\sum_{k=1}^{\infty}
X^{-+}_k \ket{k\!+\!\half^+}\bra{k^-}, \nonumber\\
&&\acute{\mm{S}}^0\mm{X}=\mm{X}\grave{\mm{S}}^0=
2\sum_{k=1}^{\infty}\bigl(w\ket{2k\!-\!1^+}\bra{2k\!-\!1^+}-w\ket{2k^-}\bra{2k^-} \nonumber \\
&&\quad - X^{++}_{2k-1}\ket{2k\!-\!\half^+}\bra{2k\!-\!\half^+}-X^{--}_{2k}\ket{2k\!-\!\half^-}\bra{2k\!-\!\half^-}\bigr) \nonumber \\
&&+2\phi
\sum_{k=0}^{\infty}
\bigl(-w \ket{2k^+}\bra{2k^+}+X^{--}_{2k+1}\ket{2k\!+\!\half^-}\bra{2k\!+\!\half^-}\bigr), \nonumber\\
&&\acute{\mm{S}}^\z\mm{X}=\mm{X}\grave{\mm{S}}^\z=
 2\sum_{k=0}^{\infty}\bigl(w \ket{2k\!+\!1^-}\bra{2k\!+\!1^-}-w\ket{2k^+}\bra{2k^+} \nonumber \\
 &&\quad + X^{++}_{2k}\ket{2k\!+\!\half^+}\bra{2k\!+\!\half^+} + X^{--}_{2k+1}\ket{2k\!+\!\half^-}\bra{2k\!+\!\half^-}\bigr) \nonumber \\
 && + 2\phi
\sum_{k=1}^{\infty}\bigl(w\ket{2k\!-\! 1^+}\bra{2k\!-\! 1^+}-X^{--}_{2k} \ket{2k\!-\!\half^-}\bra{2k\!-\!\half^-}\bigr),  \nonumber
\end{eqnarray}
and finally 
\begin{equation}
 \mm{G} \acute{\mm{S}} \mm{G} = \acute{\mm{T}},\,
 \mm{G} \grave{\mm{S}} \mm{G} = \grave{\mm{T}},\, \label{acutedef}
\end{equation}

With these definitions (i.e., Eqs. \ref{eq:Sp} -\ref{acutedef}) it was shown in \cite{PopkovProsen} that the following relations hold (recall the definition in \eqref{eq:Hhubloc}), 

\begin{eqnarray}
&& [h^{\sigma^1}_{j,j+1},\mm{S}_j \mm{X} \mm{S}_{j+1}]  = \acute{\mm{S}}_j \mm{X} \mm{S}_{j+1} - \mm{S}_j\mm{X} \grave{\mm{S}}_{j+1}, \label{eq:id1}\\
&& [h^{\sigma^2}_{j,j+1},\mm{T}_j \mm{X} \mm{T}_{j+1}] = \acute{\mm{T}}_j \mm{X} \mm{T}_{j+1} - \mm{T}_j\mm{X}\grave{\mm{T}}_{j+1}, \label{eq:id2}\\
&& \mm{S}\acute{\mm{T}} + \mm{T}\acute{\mm{S}} - \grave{\mm{S}}\mm{T} - \grave{\mm{T}}\mm{S} = [\mm{Y}-u \sigma^{1, \z} \sigma^{2,\z},\mm{S}\mm{T}], \label{eq:id3}\\
&& [\mm{S},\mm{T}] = 0, \label{eq:id4}\\
&&[\mm{X},\mm{Y}] = 0. \label{eq:id5},
\end{eqnarray}
where the subscripts in $\mm{S}_j$ ($\mm{T}_j$) indicate that they act locally on spin site $j$ of spin species 1 (2). 

Define a Lax operator $\mm{L} $ and another operator which will play the role of its "derivative" $\widetilde{\mm{L}}$,
\begin{eqnarray}
&&\mm{L} = \mm{S}\mm{T}\mm{X}, \label{eq:L} \\
&&\widetilde{\mm{L}} = \half(\mm{S}\acute{\mm{T}} + \mm{T}\acute{\mm{S}} + \grave{\mm{S}}\mm{T} + \grave{\mm{T}}\mm{S} - \{\mm{Y},\mm{S}\mm{T}\})\mm{X}, \label{eq:Lt}
\end{eqnarray}
Then using the above relations, Eqs. \ref{eq:id1} - \ref{eq:id5} and the definition in \eqref{eq:Hhubloc}, it can be shown that the so-called \emph{Sutherland-Shastry relation} holds,
\begin{equation}
[h_{j,j+1},\mm{L}_j \mm{L}_{j+1}] = (\widetilde{\mm{L}}_j + \mm{Y} \mm{L}_j)\mm{L}_{j+1} - \mm{L}_j(\widetilde{\mm{L}}_{j+1} + \mm{L}_{j+1}\mm{Y}),
\label{eq:suthsha}
\end{equation}
where $h_{j,j+1}$ is the local energy density operator given by \eqref{eq:Hhubloc1}.

Now define a central operator, 
\begin{equation}
\Omega_n := \bra{0^+} \mm{L}_1(\phi,w)\mm{L}_2(\phi,w)\cdots  \mm{L}_n(\phi,w)\ket{0^+}.
\label{eq:Om}
\end{equation}
Through this operator we may calculate a \emph{quadratically extensive} \cite{ProsenHubbardCharge} almost conserved charge, 
\begin{widetext}
\begin{eqnarray}
&&Z_1 = -\ii \partial_\varepsilon \Omega_n(\phi=0,w = \frac{\ii}{2} \varepsilon) \nonumber  \\
&&=\sum_{j=1}^{n-1}(\sigma^{1,+}_j\sigma^{1,-}_{j+1}+\sigma^{2,+}_j\sigma^{2,-}_{j+1})
- 2u\sum_{j,k}^{j<k} (-1)^{j-k}\sigma^{1,+}_j\left(\!\prod_{t=j+1}^{k-1} \sigma^{1,\z}_t\!\right) \sigma^{1,-}_k \sigma^{2,+}_j \left(\!\prod_{t=j+1}^{k-1}\sigma^{2,\z}_t\!\right)\sigma^{2,-}_k.   \label{eq:Yang2}
\end{eqnarray}
By deriving the Sutherland-Shastry relation \eqref{eq:suthsha} and using definition \eqref{eq:Om} it can be shown that this operator $Z$ is indeed almost conserved, 
\begin{equation}
[H,Z_1] = -\sigma^{1,\z}_1 - \sigma^{2,\z}_1 + \sigma^{1,\z}_n + \sigma^{2,\z}_n. \label{hub:almost1}
\end{equation}

Define an operator $S$ which is a product of spin species 2 - flip $P=\prod_{j=1}^n \sigma^{2, \rm{x}}_j$ and the transformation $u \to - u$, $S^2 = \one$.
It is obvious that $[H,S]=0$. Let, 
\begin{eqnarray}
&&Z_2=S Z_1 S= \label{yang2z2} \sum_{j=1}^{n-1}(\sigma^{1,+}_j\sigma^{1,-}_{j+1}+\sigma^{2,-}_j\sigma^{2,+}_{j+1})- 2u\sum_{j,k}^{j<k}\sigma^{1,+}_j\left(\prod_{t=j+1}^{k-1} \sigma^{1,\z}_t \right) \sigma^{1,-}_k \sigma^{2,-}_j \left(\prod_{t=j+1}^{k-1}\sigma^{2,\z}_t\right)\sigma^{2,+}_k, \nonumber 
\end{eqnarray}
\end{widetext}
 and then by multiplying \eqref{hub:almost1} from the left and the right by $S$ we find,
\begin{equation}
[H,Z_2] = -\sigma^{1,\z}_1 + \sigma^{2,\z}_1 + \sigma^{1,\z}_n - \sigma^{2,\z}_n. \label{hub:almost2}
\end{equation}
This completes the review of the work \cite{PopkovProsen} we need to continue and now we can finally state and solve our problem. We will study driving of the Hubbard with a total of eight Lindblad operators, 
\begin{eqnarray}
L_{d_+}^{1,+}=\sigma^{d, +}_1 &\quad& L_{d_-}^{1,+}=\sigma^{d -}_1 \\
L_{d_+}^{n,-}=\sigma^{d, +}_n &\quad& L_{d_-}^{1,-}=\sigma^{d -}_n,
\end{eqnarray}
where $d=1,2$ designates either spin species 1 or 2 with appropriate driving parameters $\mu_{1,2}$. The rest of the setup is defined in the previous section and the perturbation expansion is the same as before \eqref{eq:pert}. 

The Hubbard Hamiltonian fulfils the parity symmetry requirement \eqref{eq:Psym2_hub} with $P=\prod_{d=1}^2\prod_{j=1}^n \sigma^{d, \rm{x}}_j$ being a global spin flip operator in the $\rm{z}$-basis.

Clearly, the zeroth order solution is $\rho^{(0)}=4^{-n} \one $ and the first order solution is given through the conserved charges in \eqref{eq:Yang2}. By applying the total dissipator $\DD_{\vec{\chi}} \rho^{(0)}$ and using the algebraic condition \eqref{eq:cond_hub} we find that the solution to the first order equation \eqref{fo} is,
\be
\rho^{(1)}=\sum_{d=1}^2 c^{(1)}_{d}( Z_d-Z_d^{\dagger}), \label{fos_hub} 
\ee
where,
\begin{eqnarray}
&c^{(1)}_1=\frac{1}{4} (\mu_1 \left( \cos \chi_1+1\right)-\mu_2 \left(\cos \chi_2+1\right) \\
&-\ii \left(\sin \chi _1+\sin \chi _2\right)), \nonumber \\
&c^{(1)}_2= \frac{1}{4} (\mu_1 \left( \cos \chi_1+1\right)+\mu_2 \left(\cos \chi_2+1\right) \\
&-\ii \left(\sin \chi _1+\sin \chi _2\right)). \nonumber \end{eqnarray}
Moving on to the second order equation,
\be
 (\!\ad H)\rho^{(2)} + \hat{\cal D}_{\vec{\chi}} \rho^{(1)}= \lambda^{(1)} \rho^{(1)},
 \label{eq:seco}
\ee
we recall that like for the $XXZ$ spin chain  \cite{Prosen1} $Z_{1,2}$ contain no terms with factors $\sigma_j^{d,\z}$ ($d=1,2$). We then only need to check \eqref{eq:seco} for $Z_{d} \in\{ \one, \sigma^{d, +}_{1,n}, \sigma^{d, -}_{1,n} \}$. We take a quadratic ansatz for the second order solution,
\be
\rho^{(2)}=\sum_{j,k} (c^{(2)}_{j,k} Z_j Z_k+c^{(2)}_{j^\dagger,k} Z^\dagger_j Z_k+ c^{(2)}_{j,k^\dagger} Z_j Z^\dagger_k +c^{(2)}_{j^\dagger,k^\dagger} Z^\dagger_j Z^\dagger_k). \label{rho2hub}
\ee
The calculation for the second-order coefficients is now straightforward, but tedious, and we can find the expressions for the coefficients $c^{(2)}_{j,k}, c^{(2)}_{j^\dagger,k},  c^{(2)}_{j,k^\dagger},c^{(2)}_{j^\dagger,k^\dagger}.$ They are quite long and not directly relevant for the rest of the discussion so we omit them here and instead will write the explicit expression in Appendix ~\ref{app:Hubbard}. 

\begin{figure}
 \centering	
\vspace{-1mm}
\includegraphics[width=0.99\columnwidth]{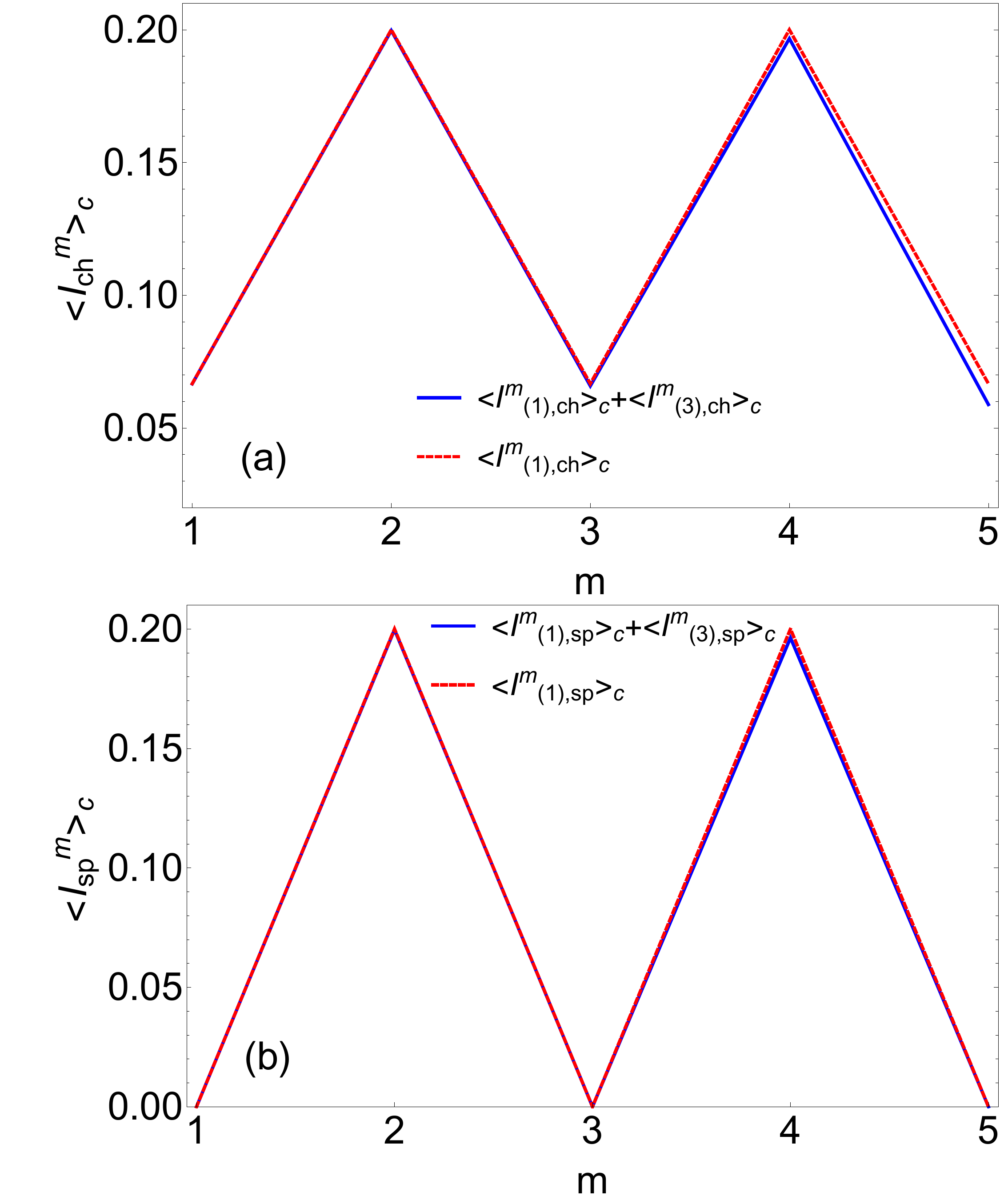}
\vspace{-1mm}
\caption{The first five (a) charge and (b) spin current cumulants for $\mu _1=\mu _2=1/3$ at weak system-bath coupling $\varepsilon=0.05$ and intermediate interaction strength $u=2$. Note that the choice of $\mu _1=\mu _2$ corresponds physically to the Lindblad jump operators taking in and putting out the same number of spin-up and spin-down fermions on average, which is reflected in the fact that the odd cumulants are zero in (b). The red (dashed) lines are the first order corrections and the blue (full) ones are the third order ones.}
\label{fig:hub2}
\end{figure}

Like before this determines $\rho^{(2)}$ up to the addition of conserved quantities $Q_k$, where $[H,Q_k]=0$, that is, $\rho^{(2)'} =  \rho^{(2)} + \sum_k \alpha_k Q_k$. However, like in \cite{countstatweak} they are irrelevant because via the third order equation, 
 \be
(\!\ad H) \rho^{(3)} + \hat{\cal D}_{\vec{\chi}} \rho^{(2)'}=  \lambda^{(3)} \rho^{(0)}+\lambda^{(1)} \rho^{(2)'},
\ee
we find that,
 \be
 \lambda^{(3)} = \tr (\hat{\cal D}_{\vec{\chi}} \rho^{(2)'}- \lambda^{(1)} \rho^{(2)'}), \label{tr3_hub}
\ee
 and due to the form of the dissipator only terms of the form $O \in \{\one, (-\sigma^{d,\z}_1+\sigma^{d,\z}_n) \}$ in $\rho^{(2)'}$ can contribute to \eqref{tr3_hub}. Now since $ (-\sigma^{d, \z}_1+\sigma^{d, \z}_n) \in  {\rm Im}\ad H $ for $d=1,2$ due to the existence of a solution to the first order equation \eqref{fo} merely $\rho^{(2)}$ is sufficient to determine the third order correction,
 \be
 \lambda^{(3)} =\sum_{d=1}^2 (-\mu_d-\mu_d \cos\chi_d + \ii \sin\chi_d)\tr[\bigl(-\sigma^{d, \z}_1+\sigma^{d,\z}_n) \rho^{(2)}\bigr] \label{tr3ah}. 
\ee
The full form of $ \lambda^{(3)}$ is long and is given as Eq.~\eqref{lambda3} in the appendix. Now let us finally turn to computing the third-order correction to the current cumulants. Note that there are two physically relevant currents we can study, first the charge current (corresponding to setting $\chi=\chi_1= \chi_2\ $) and secondly the spin current (corresponding to setting $\chi=\chi_1= -\chi_2\ $). 
In the first case we have for charge current for odd and even $n$, 
\begin{widetext}
\begin{eqnarray}
&&\left <I^{2k-1}_{(3),\rm{ch}} \right >_{\rm{c}}=  -\varepsilon^3\frac{\left(\mu _1+\mu _2\right) \left(\left(9^k-9\right) \left(-8 \mu _2 \mu _1+\mu _1^2 \left(u^2+8\right)+\mu _2^2 \left(u^2+8\right)\right)+6 \left(9^k-1\right) \left(u^2+4\right)\right)}{384 (2 k-1)!} \nonumber \\
&&\left <I^{2k}_{(3),\rm{ch}} \right >_{\rm{c}}=-\varepsilon^3 \frac{\left(9^k-1\right) \left(\mu _1 \mu _2 u^2+\mu _1^2 \left(u^2+6\right)+\mu _2^2 \left(u^2+6\right)+u^2+4\right)}{32 (2 k)!}. \label{hubcums}
\end{eqnarray}
\end{widetext}
The spin current cumulants $\left <I^{n}_{(3),\rm{sp}} \right >_{\rm{c}}$ can be obtained from the charge current ones simply by setting $\mu_1 \to -\mu_1$ in \eqref{hubcums}. We plot them in Fig. ~\ref{fig:hub2}. 

It is interesting to note that if we set $\mu=\mu _1=\mu _2$ the charge current cumulants become quite similar to the ones for the $XXZ$ spin chain \cite{countstatweak}. They only differ up to a constant replacing $u$ in \eqref{hubcums}. This reflects potentially a deep symmetry connection between the two models. Indeed, since the derivation only depended on the fact that  $Z_{1,2}$ contain only terms of the form $Z_{d} \in\{ \one, \sigma^{d, +}_{1,n}, \sigma^{d, -}_{1,n} \}$ on sites $1$ and $n$ this could hint at a possible universal result for this class of integrable models. 

We can also calculate the charge and spin current connected correlations using the explicit expression for $\lambda^{(1)}$ and $\lambda^{(3)}$ (eqs. \eqref{zero_hub_eig}, and \eqref{lambda3} in the appendix, respectively) by using a generalization of \eqref{cs0_hub} for multiple counting fields,
\begin{align}
&\ave{I_{d_1}^{m_1}I_{d_2}^{m_2}I_{d_3}^{m_3} \ldots I_{d_p}^{m_p} }_c \\ 
&=  \frac {\partial^{m_1+m_2+m_3 \ldots m_p} \lambda(\vec{\chi}) }{\partial (\ii \chi_{d_1})^{m_1} \partial (\ii \chi_{d_2})^{m_2}\partial (\ii \chi_{d_3})^{m_3} \ldots \partial (\ii \chi_{d_p})^{m_p}} \Big |_{\vec{\chi} \to 0}, \nonumber
\end{align}
where in the case of two counting fields $d_k=1,2$ and we will again transform into the spin-charge picture using $\chi_1=(\chi_{\rm{ch}}+\chi_{\rm{sp}})/2$ and $\chi_2=(\chi_{\rm{ch}}-\chi_{\rm{sp}})/2$.
The first few are, 
\begin{align}
&\left <I_{\rm{ch}} I_{\rm{sp}} \right >_{\rm{c}}= \varepsilon^3 \frac{1}{16} \left(\mu _1^2-\mu _2^2\right) \left(u^2+6\right) +\mathcal{O}(\varepsilon^5),  \\
&\left <I^2_{\rm{ch}} I_{\rm{sp}} \right >_{\rm{c}}= \frac{\varepsilon}{4} \left(\mu _1-\mu _2\right) \\ 
&+\frac{\varepsilon^3}{128}  \left(\mu _1-\mu _2\right) [3 \mu _1^2 \left(u^2+8\right)+3 \mu _2^2 \left(u^2+8\right) \nonumber \\
&+4 \mu _1 \mu _2 \left(u^2+6\right)+12 u^2+80]+\mathcal{O}(\varepsilon^5), \nonumber \\
\end{align}
and,
\begin{align}
&\left <I_{\rm{ch}} I^2_{\rm{sp}} \right >_{\rm{c}}=  \frac{\varepsilon}{4} \left(\mu _1+	\mu _2\right) \\ 
&+\frac{\varepsilon^3}{128}  \left(\mu _1+\mu _2\right) [3 \mu _1^2 \left(u^2+8\right)+3 \mu _2^2 \left(u^2+8\right) \nonumber \\
&-4 \mu _1 \mu _2 \left(u^2+6\right)+12 u^2+80]+\mathcal{O}(\varepsilon^5), \nonumber \\
&\left <I^2_{\rm{ch}} I^2_{\rm{sp}} \right >_{\rm{c}}=  \frac{\varepsilon}{4}  \\ 
&+\frac{\varepsilon^3}{32}  \left(-2 \mu _1^2 \left(2 u^2+15\right)-2 \mu _2^2 \left(2 u^2+15\right)-3 u^2-20\right)\nonumber \\
&+\mathcal{O}(\varepsilon^5). \nonumber
\end{align}
The general expressions for arbitrary order correlations between two currents ate given in appendix. One can also calculate higher order connected correlations from the expression for  $\lambda^{(1)}$ and $\lambda^{(3)}$ (eqs. \eqref{zero_hub_eig} and \eqref{lambda3} in the appendix, respectively, give the two-point connected correlation functions).  
\section{Conclusion}

In this paper we generalized the results of Ref.~\cite{countstatweak} for the full counting statistics of parity symmetric spin systems in the weak-system bath coupling regime to multiple baths and spin species in the leading order. Then later we calculated the the next-to-leading order for the charge and spin current statistics for the case of the Hubbard model in the next-to-leading order, as well as the charge-spin correlations. The form of the charge current cumulants are similar to those in the $XXZ$ case \cite{countstatweak} indicating a possible universal feature of these class of integrable models. Based on recent work \cite{LemosProsen} likely do not hold in the thermodynamic limit due to issues with this perturbation expansion, however they should be applicable to finite size system. One would thus desire a fully non-perturbative extension for the $XXZ$ and the Hubbard model, which currently remain elusive.

\bigskip
\bigskip

\section[Appendix:The second-order correction to the FCS \ldots]{Appendix: The second-order correction to the full counting statistics of the open Hubbard model}
\label{app:Hubbard}

\subsection{Explicit coefficients of the second order correction to the counting eigenvector}

Here we will write out the coefficient in the second-order correction to the eigenvector corresponding to the leading eigenvalue of the deformed Liouvillian of the open Hubbard model Eq. \eqref{rho2hub} and the full form of the third-order correction to the current cumulant generating function. 

The coefficients multiplying operators which are not conjugate transposed are,
\begin{widetext}
\begin{align}
&c_{1,2}= \nonumber \\
&\frac{1}{4} \cos \left(\frac{\chi _2}{2}\right) \left(\mu _2 \cos \left(\frac{\chi _2}{2}\right)+i \sin \left(\frac{\chi _2}{2}\right)\right) \left(-\ii \mu _1 \sin\chi _1+\ii \mu _2 \sin\chi _2-\cos\chi _1+\cos\chi _2\right)\\
&c_{2,1}=\nonumber \\
& \frac{1}{64}\Big\{24 \mu _1^2 \left(U^2+2\right) \cos ^4\left(\frac{\chi _1}{2}\right)+4 \mu _2 \cos ^2\left(\frac{\chi _2}{2}\right) [\mu _2 \{\left(3 U^2+2\right) \left(\cos\chi _2+1\right)\nonumber \\
&-4 i \sin\chi _2\}+2 i \left(3 U^2+2\right) \sin\chi _2+4 \cos\chi _1-8 \cos\chi _2+4]\nonumber\\
&+4 \ii \mu _1 \sin\chi _1 \{2 \mu _2 \left(\cos\chi _2+1\right)\nonumber\\
&+3 \left(U^2+2\right) \left(\cos\chi _1+1\right)+2 i \sin\chi _2\}+3 \left(U^2+2\right) \cos \left(2 \chi _1\right)+\left(3 U^2+2\right) \cos \left(2 \chi _2\right)\nonumber \\
&-6 U^2-4 i \sin \left(2 \chi _2\right)+8 i \sin\chi _2 \cos\chi _1-8 \Big\}
\end{align}

The mixed ones are,

\begin{align}
&c_{1^\dagger,1}=\frac{1}{16} \left(\mu _1 \left(1+e^{i \chi _1}\right)+\left(\mu _2+1\right) e^{i \chi _2}+\mu _2+e^{i \chi _1}\right) \{\mu _1 \left(\cos\chi _1+1\right)+\mu _2 \left(\cos\chi _2+1\right)\nonumber\\
&+\ii \left(\sin\chi _1+\sin\chi _2\right)\}\\
&c_{1,1^\dagger}= \frac{1}{16}\Big\{-\frac{1}{2} \mu _1^2 e^{-2 i \chi _1} \left(1+e^{i \chi _1}\right)^3+\mu _2 e^{-\frac{1}{2} \left(i \chi _2\right)} \cos \left(\frac{\chi _2}{2}\right)\nonumber\\
&\times \{-2 \mu _2 \left(\cos\chi _2+1\right)-i \left(3 \sin\chi _1+\sin \left(\chi _1-\chi _2\right)+3 \sin\chi _2+i\right)+\cos\chi _1\nonumber\\
&+\cos \left(\chi _1-\chi _2\right)+\cos\chi _2\}+\mu _1 e^{-\frac{1}{2} i \left(\chi _1+2 \chi _2\right)} \cos \left(\frac{\chi _1}{2}\right) [\left(\mu _2-1\right) \left(-2-e^{i \chi _1}\right)\nonumber\\
&-\left(\mu _2+1\right) e^{2 i \chi _2}+e^{i \chi _2} \left(-\mu _2 \left(3+e^{i \chi _1}\right)-3 i \sin\chi _1+\cos\chi _1+1\right)]+\{\sin\chi _1\nonumber\\
&+\sin\chi _2\} \left(\sin\chi _1+\sin\chi _2+i \cos\chi _1+i \cos\chi _2\right) \Big\} \\
&c_{2^\dagger,2}=\nonumber\\
&\frac{1}{16} e^{-i \chi _2} \left(e^{i \chi _2} \left(\left(\mu _1+1\right) e^{i \chi _1}+\mu _1-\mu _2\right)-\mu _2+1\right) \{-\mu _1 \left(\cos\chi _1+1\right)+\mu _2 \left(\cos\chi _2+1\right)\nonumber\\
&-i \left(\sin\chi _1-\sin\chi _2\right)\}\\
&c_{2,2^\dagger}=\frac{1}{16} e^{-i \chi _1} \left(\mu _1 \left(-\left(1+e^{i \chi _1}\right)\right)+\mu _2 e^{i \chi _1}+\left(\mu _2+1\right) e^{i \left(\chi _1+\chi _2\right)}+1\right)\nonumber\\
 &\times\left(\mu _1 \left(\cos\chi _1+1\right)-\mu _2 \left(\cos\chi _2+1\right)+i \left(\sin\chi _1-\sin\chi _2\right)\right),
\end{align} 

and,

\begin{align}
&c_{1,2^\dagger}=\frac{1}{16} e^{-i \chi _1} \left(\mu _1 \left(1+e^{i \chi _1}\right)+\mu _2 e^{i \chi _1}+\left(\mu _2+1\right) e^{i \left(\chi _1+\chi _2\right)}-1\right)\nonumber \\
&\times\left(-\mu _1 \left(\cos\chi _1+1\right)+\mu _2 \left(\cos\chi _2+1\right)-\ii \left(\sin\chi _1-\sin\chi _2\right)\right)\\
&c_{2^\dagger,1}=\frac{1}{64}\Big\{ 8 \mu _2^2 \cos ^3\left(\frac{\chi _2}{2}\right) \left(\left(3 U^2+4\right) \cos \left(\frac{\chi _2}{2}\right)-2 i \sin \left(\frac{\chi _2}{2}\right)\right)\nonumber \\
&+4 \mu _1 \{\mu _1 \cos ^2\left(\frac{\chi _1}{2}\right) \left(\left(3 U^2+4\right) \left(\cos\chi _1+1\right)-2 i \sin\chi _1\right)\nonumber\\
&+2 \cos ^2\left(\frac{\chi _1}{2}\right) \left(\ii \left(3 U^2+4\right) \sin\chi _1-2 \cos\chi _1+\cos\chi _2+1\right)-\sin\chi _1 \sin\chi _2\}\nonumber\\
&+4 \mu _2 [\ii \mu _1 \left(\sin\chi _1+\sin\chi _2+\sin \left(\chi _1+\chi _2\right)\right)+\cos ^2\left(\frac{\chi _2}{2}\right) \nonumber\\
&\times \left(2 \ii \left(3 U^2+4\right) \sin\chi _2-4 \cos\chi _2+2\right)\nonumber\\
&+\cos\chi _1+\cos \left(\chi _1+\chi _2\right)]-2 \left(3 U^2+i \sin \left(2 \chi _1\right)+i \sin \left(2 \chi _2\right)-2 i \sin \left(\chi _1+\chi _2\right)+4\right)\nonumber\\
&+\left(3 U^2+4\right) \cos \left(2 \chi _1\right)+\left(3 U^2+4\right) \cos \left(2 \chi _2\right) \Big\}.
\end{align}

Continuing,

\begin{align}
&c_{1^\dagger,2}=\frac{1}{16} e^{-i \chi _1} \left(\mu _1 \left(-\left(1+e^{i \chi _1}\right)\right)+\mu _2 e^{i \chi _1}+\left(\mu _2+1\right) e^{i \left(\chi _1+\chi _2\right)}+1\right)\nonumber\\
 &\times\left(\mu _1 \left(\cos\chi _1+1\right)+\mu _2 \left(\cos\chi _2+1\right)+i \left(\sin\chi _1-\sin\chi _2\right)\right),\\
& 64c_{2,1^\dagger}=8 \mu _2^2 \cos ^3\left(\frac{\chi _2}{2}\right) \left(\left(3 U^2+4\right) \cos \left(\frac{\chi _2}{2}\right)- i \sin \left(\frac{\chi _2}{2}\right)\right)\nonumber \\
&+2 \mu _1 \{\mu _1 \cos ^2\left(\frac{\chi _1}{2}\right) \left(\left(3 U^2+4\right) \left(\cos\chi _1+1\right)- i \sin\chi _1\right)\nonumber\\
&+2 \cos ^2\left(\frac{\chi _1}{2}\right) \left(i \left(3 U^2+4\right) \sin\chi _1-2 \cos\chi _1+\cos\chi _2+1\right)-\sin\chi _1 \sin\chi _2\},
\end{align}
\end{widetext}
while the rest of coefficients are zero. Note that several pairs are similar in form, though not exactly the same.

The full form of the third-order correction to the cumulant generating function is given as,
\begin{widetext}
\begin{eqnarray}
 &&\lambda^{(3)} =\frac{1}{64} \left(\ii \mu _1 \sin \chi _1+\ii \mu _2 \sin \chi _2+\cos \chi _1+\cos \chi _2\right) [-2 \mu _1^2 \left(u^2+8\right) \sin ^2\chi _1  \label{lambda3}\\
 &&+4 \ii \mu _1 \sin \chi _1 \left(-4 i \mu _2 \sin \chi _2+\left(u^2+8\right) \cos \chi _2-4 \cos \chi _2\right) \nonumber \\
 &&-2 \mu _2 \sin \chi _2 \left(\left(u^2+8\right) \left(\mu _2 \sin \chi _2-2 \ii \chi _1\ii \cos \chi _2\right)+8 i \cos \chi _1\right) \nonumber \\
 &&+\left(u^2+8\right) \cos \left(2 \chi _1\right)-2 \left(u^2+8 \cos \chi _1 \cos \chi _2\right)+\left(u^2+8\right) \cos \left(2 \chi _2\right)]. \nonumber
\end{eqnarray}

\subsection{General spin-charge current correlations in the driven Hubbard model}

We finally give the full form of the spin-charge current correlations. First the odd-odd ones,

\begin{eqnarray}
&&\left <I^{2k-1}_{(3),\rm{ch}} I^{2m-1}_{(3),\rm{sp}} \right >_{\rm{c}}= \nonumber \\
&& -\varepsilon^3\frac{\left(\mu _1^2-\mu _2^2\right) \left(u^2 \left(9^{k+m}+9^k+9^m-3\right)+72 \left(9^k-1\right)\right)}{256 (2 k-1)!(2 m-1)!} ,\nonumber
\end{eqnarray}
then the even-odd ones, 
\begin{eqnarray}
&&\left <I^{2k-1}_{(3),\rm{ch}} I^{2m}_{(3),\rm{sp}} \right >_{\rm{c}}=\nonumber\\
&&-\varepsilon^3 \frac{\left(\mu _1+\mu _2\right) \left(\mu _1^2 \left(u^2+8\right) \left(9^{k+m}-9\right)-\mu _2 \mu _1 \left(\left(9^k+3\right) \left(9^m-1\right) u^2+8 \left(9^{k+m}-9\right)\right)\right)}{32 (2 k-1)! (2 m)!} \nonumber \\
 &&-\varepsilon^3 \frac{\left(\mu _1+\mu _2\right) \left(\mu _2^2 \left(u^2+8\right) \left(9^{k+m}-9\right)+3 \left(u^2 \left(9^{k+m}+9^k+9^m-3\right)+8
   \left(9^{k+m}-1\right)\right)\right)}{64 (2 k-1)! (2 m)!},
\end{eqnarray}
the odd-even ones can be obtained by setting $\mu_2 \to -\mu_2$ and interchanging $m$ and $k$, and the even-even ones are,
\begin{eqnarray}
&&\left <I^{2k}_{(3),\rm{ch}} I^{2m}_{(3),\rm{sp}} \right >_{\rm{c}}=\nonumber\\
&&-\varepsilon^3 \frac{\mu _1^2 \left(u^2 \left(3^{2 k+2 m+1}+9^k+9^m-5\right)+24 \left(9^{k+m}-1\right)\right)+2 \left(\left(u^2+8\right) 9^{k+m}+2\ 9^k u^2-3 u^2-8\right)}{32 (2 k)! (2 m)!} \nonumber \\
 &&-\varepsilon^3 \frac{4 \mu _1 \mu _2 u^2 \left(9^k-9^m\right)+\mu _2^2 \left(u^2 \left(3^{2 k+2 m+1}+9^k+9^m-5\right)+24 \left(9^{k+m}-1\right)\right)}{32 (2 k)! (2 m)!}.
\end{eqnarray}
\end{widetext}


\begin{thebibliography}{10}
\bibitem{footnote} We use this notation for basis vectors because the Lax operator for the open maximally driven Hubbard model can also be written in terms of a very elegant \emph{walking graph state representation} \cite{ProsenHubbard}. Although we do not address that in this paper we retain this labelling for sake of consistency with previous literature to which we refer the reader.
\end{thebibliography}
\end{document}